# 50/500 or 100/1000 debate is not about the time frame – Reply to Rosenfeld

Richard Frankham

*Department of Biological Sciences, Macquarie University, NSW 2109, Australia*
Tel.: +61 2 9850 8186; fax: +61 2 9850 8245.
E-mail address: richard.frankham@mq.ed.au

Corey J.A. Bradshaw and Barry W. Brook

*The Environment Institute and School of Earth and Environmental Sciences, The University of Adelaide, South Australia 5005, Australia*
E-mail addresses: corey.bradshaw@adelaide.edu.au and barry.brook@adelaide.edu.au

--

The Letter from Rosenfeld (2014) in response to Jamieson and Allendorf (2012) and Frankham et al. (2014) and related papers is misleading in places and requires clarification and correction, as follows:

1. "Census population size (the MVP) may be anywhere from 5 to 10 times the effective population size": this is far too low for highly fecund species, such as fish where it is typically 1000 or more (Palstra & Ruzzante 2008; Frankham et al. 2014).

2. "Frankham and associates … goals focus on recovery targets needed for long-term persistence in perpetuity, rather than short-term prevention of extinction…": Incorrect: the $N_e$ = 50 (or 100) that is a major focus of our paper (including in the title) deals explicitly with avoiding inbreeding depression in the short-term and thus avoiding immediate extinctions. Furthermore, the five-generations duration we recommend is most definitively 'short-term'. Frankham and colleagues have also done many laboratory and modelling experiments on inbreeding and extinction and reviewed the field several times, concluding that the inbreeding avoidance we recommend applies to short-term management time frames (see Frankham 2005; Frankham et al. 2010 for references).

3. "Frankham et al. (2014) do not appear to dispute the case made by Jamieson and Allendorf (2012) that most endangered species are rapidly declining as a consequence of human impacts … that increase mortality and that the effects of inbreeding and reduced evolutionary potential are secondary.": this is misleading; some are, but typically it is a synergy of deterministic and stochastic effects (including genetic ones) that cannot be arbitrarily disentangled, as established by several independent approaches (described and referenced in Frankham et al. 2014). First, population viability analyses for many real, threatened species revealed that inclusion of inbreeding depression in stochastic-demographic models resulted in median reductions of 30-40% in median times to extinction – such reductions are clearly not "secondary" (Brook et al. 2002; O'Grady et al. 2006). Second, the related view that other factors typically drive species to extinction before genetic factors can impact them has been refuted, based on > 170 comparisons of genetic diversity in threatened and closely related non-threatened species (Spielman et al. 2004; Evans & Sheldon 2008; Flight 2010). Third, empirical field studies, where relative contributions could be partitioned, have also demonstrated large effects of inbreeding and loss of genetic diversity on extinction risk (Newman & Pilson 1997; Saccheri et al. 1998; Nieminen et al. 2001; Vilas et al. 2006). Fourth, gene flow into inbred populations with low genetic diversity typically leads to large genetic rescue effects on fitness, especially in natural outbreeding species (Tallmon et al. 2004; Frankham, unpublished data). Fifth, small populations of self-incompatible species with so few S alleles that they are functionally extinct become capable of sexual reproduction when outcrossing adds new S alleles, as observed in the Illinois population of the Lakeside daisy and in Florida ziziphus (DeMauro 1993; Weekley et al. 2002; Gitzendanner et al. 2012).

4. "… much of this debate concerns the time scale over which to plan for recovery and persistence … rather than the science": while there are substantial areas of agreement between the groups, Jamieson and Allendorf disagree with our revision from 50/500 to 100/1000 (Franklin et al. 2014).

5. "However, long-term PVAs that account for evolutionary potential are not at all incompatible with PVAs over shorter time horizons that identify priority threats to persistence and interim recovery targets." We did not say otherwise.

6. Neither our paper, nor our previous work related to the issue (e.g., Traill et al. 2010) "… implicitly link long-term MVPs in the thousands to a triage approach that may explicitly write off extremely rare species that are costly to recover", as suggested by Jamieson and Allendorf (2012) and repeated by Rosenfeld (2014). Frankham et al. (2014) did not discuss triage, which is essentially "smart decision making" (Bottrill et al. 2008).

7. Finally, framing the question of short- or long-term management of extinction risk as mutually exclusive goals is moot if the ultimate outcome is extinction. Separating the two time frames is therefore fallacious.